\documentclass{elsart}

 \usepackage{epsfig}

\usepackage{amssymb}
\usepackage{amsmath}
\begin{document}

\begin{frontmatter}



\title{Deterministic Scale-Free Networks}

 \author{Albert-L\'aszl\'o Barab\'asi$^a$, Erzs\'ebet Ravasz$^a$, Tam\'as Vicsek$^b$}
 \address{$^a$Department of Physics, University of Notre Dame, \\ Notre Dame, IN 46556-5670, USA\\
	  $^b$Department of Biological Physics, E\"otv\"os University, \\ 
		P\'azm\'any P\'eter S\'etany	1-A, Budapest, Hungary, H-1117}

\begin{abstract}

    Scale-free networks are abundant in nature, describing such diverse
systems as the world wide web, the web of human sexual contacts, or the chemical network of a
cell.
    All models used to generate a scale-free topology are stochastic,
that is they create networks in which the nodes appear to be randomly
connected to each other.
    Here we propose a simple model that
generates scale-free networks in a deterministic  fashion.
    We solve exactly the model, showing that the tail of the
degree distribution follows a power law.

\end{abstract}

\end{frontmatter}


    Networks are ubiquitous in nature and society \cite{reka,alb,dorog},
describing various complex systems, such as the society, a
network of individuals linked by various social links \cite{amaral,newman1,newman2,collab};
the Internet, a network of routers connected by various physical
  connections \cite{falaustos,attack};
the world wide web, a virtual web of documents connected by uniform
resource connectors \cite{diameter} or the cell, a network
of substrates connected by chemical reactions  \cite{43_bio,protein}.
     Despite their diversity, most networks appearing in nature follow universal organizing principles.
    In particular, it has been found that many
     networks of scientific interest are scale-free \cite{BA_sci,BA_phys}, that is, the probability that a randomly selected node has exactly
      $k$ links decays as a power law, following $P(k) \sim k^{-\gamma}$, where $\gamma$ is the degree exponent.
    The list of documented
scale-free networks now include the world wide web \cite{diameter,broder}, the Internet \cite{falaustos},
the cell \cite{43_bio,protein}, the web of human sexual contacts \cite{sex}, the language  \cite{sole}, or the
web of actors in Hollywood \cite{actor}, most of which appear to have degree exponents between two and three.

    The $\,$high$\,$ interest$\,$ in$\,$ understanding$\,$ the$\,$
 topology$\,$ of $\,$complex $\,$networks$\,$
has $\,$resulted$\,$ in$\,$ the$\,$ development$\,$ of a$\,$ considerable$\,$ number $\,$of $\,$
network $\,$models \cite{BA_sci,BA_phys,m1,m2,m3,m4,r1,r2,r3,gin1,gin2}.
    Most of these are based on two mechanisms: incremental growth and preferential attachment \cite{BA_sci,BA_phys}.
    Incremental  growth captures the fact that  networks are assembled
through the addition of new nodes to the system, while preferential
 attachment encodes the hypothesis that new nodes connect with higher probability
to more connected nodes.
    Both of these mechanisms have been supported by extensive
empirical measurements \cite{newman2,zoli,pastor}, indicating that they are
 simultaneously present in many systems with scale-free network topology.

    Stochasticity is a common feature of all network models that generate
 scale-free topologies.
    That is, new nodes connect using
a probabilistic rule to the nodes already present in the system.
    This randomness present in the models, while in line with
the major features of networks seen in nature, makes it harder to gain a visual understanding of what makes them
scale-free, and how do different
nodes relate to each other.
    It would therefore be of major theoretical
interest to construct models that lead to scale-free
networks in a deterministic fashion.
    Here we present such a simple model, generating a
deterministic scale-free network using a hierarchical
construction.

\section{Model description}
\label{model}
    The construction of the model, that follows a hierarchical rule commonly used in
deterministic fractals \cite{mandel,vicsek}, is shown in Figure 1.
    The network is built in an iterative fashion, each iteration
repeating and reusing the elements generated in the previous
steps as follows:
\begin{figure}[h]
\begin{center}
\epsfig{figure=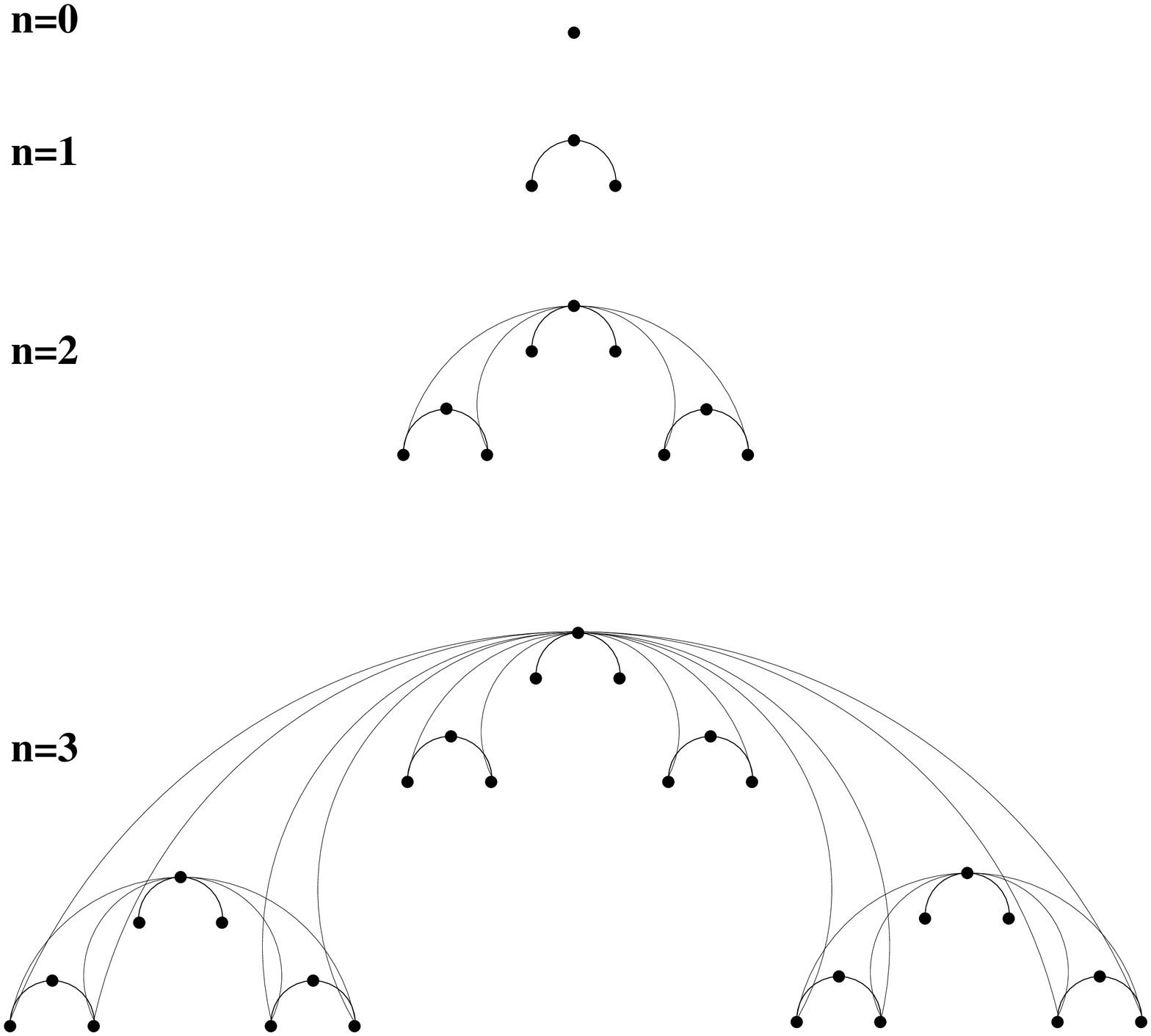,height=8.5cm,width=9.8cm}
\caption{Construction of the deterministic scale-free network,
showing the first four steps of the iterative process.}
\label{fig:fig1}
\end{center}
\end{figure}

\emph{Step 0:} We start from a single node, that we designate as the 
\emph{root} of the graph.

\emph{Step 1:} We add two more nodes, and connect each of them to the root.

\emph{Step 2:} We add two units of three nodes, each unit identical
to the network created in the previous iteration (step 1), and we
 connect each of the {\it bottom}  nodes (see Figure 1) of these two units to the
 root.
 That is, the root will gain four more new links.

\emph{Step 3:} We add two units of nine nodes each, identical
to the units generated in the previous iteration, and connect all
eight bottom nodes of the two new units to the root.

These rules can be easily generalized. Indeed,  step $n$ would involve the following operation:

\emph{Step $n$:} Add two units of $3^{n-1}$ nodes each, identical
to the network created in the previous iteration (step $n-1$), and
connect each of the $2^n$ bottom nodes of these two units to the
root of the network.

\section{Analytical solution}
\label{anal}

Thanks to its deterministic and discrete nature, the model
described above can be solved exactly. To show its scale-free
nature, in the following we concentrate on the degree
distribution, $P(k)$.

    The tail of the degree distribution is determined by the most connected nodes, or hubs.
    Clearly the biggest hub is the root, and the next two hubs are the roots of the two
    units added to the network in the last step.
    Therefore, in order to capture the tail of the distribution, it is sufficient to focus
     on the hubs.

    In step $i$ the degree of the most connected hub, the root, is $2^{i+1}-2$.
     In the next iteration two
    copies of this
hub will appear in the two newly added units. As we iterate
further, in the $n$th step $3^{n-i}$ copies of this hub will be
present in the network. However,  the two newly created copies
will not increase their degree after further iterations.
  Therefore, after $n$ iterations  there are $(2/3) 3^{n-i}$
  nodes with degree  $2^{i+1}-2$. Since spaces between degrees grow with increrasing $k$,
the exponent of the degree distribution can be calculated using the cumulative degree 
disctribution. The tail of the cumulative degree distribution, determined by the hubs, follows
$$
    P_{\text{cum}}(k) \sim k^{1-\gamma} \sim k^{-\frac{\ln3}{\ln2}}.
$$
 Thus the degree exponent is $$\gamma=1+\frac{\ln3}{\ln2}.$$
    The origin of this scaling behavior can be understood by inspecting the model's construction.
    Indeed, at any moment we have a hierarchy of hubs, highly
connected nodes which are a common component of scale-free
networks.  The root is always the largest hub.
    However, at any step there are two hubs whose connectivity is roughly a half of the root's
connectivity, corresponding to the roots of the two units added at step $n-1$.
    There are six even smaller hubs, with connectivity $2^{n-1}-2$, corresponding to
the root of the units added at time $n-2$, and so on. This
hierarchy of hubs is responsible for the network's scale-free
topology.

\section{Discussion}
\label{disc}
The model introduced above offers a deterministic construction of a scale-free network.
    One of its interesting properties is its apparent
self-similarity.
    On the other hand, while the networks generated at step $n$ and $n-1$ have an identical
large-scale topology, self-similarity is not complete.
    The special role played by the root,
the major hub in the network, violates complete self-similarity.
    Indeed, the root keeps a detailed record of the system size through the
number of links it has.
    Such global information is never present in a local element of a
fractal, the common example of a self-similar object.
    This unique feature of our construction reflects a common
property of all scale-free networks, rooted in the nonlocal growth
rule that generates scale-free systems.
       Indeed, in stochastic models the probability that a node
connects to an existing node in the system contains implicitly
the information about the whole system.
    While the connectivity of the hub in a stochastic
network is not equal to the system size, it typically varies as a simple function of $N$.
    Similarly, here the largest hub, the root of our construction, keeps track of the
       system size in a trivial way, as $2/3$ of the nodes are linked to it.

The proposed model generates a network with a fixed
$\gamma=1+\ln3/\ln2$ degree exponent.
    However, one can easily modify the model to change
the scaling exponent by varying the number of links connected to
the root at each step. 
    Similarly,  by definition the model discussed
here has a zero clustering coefficient  \cite{newman2}, as it does
not generate triangles of connected nodes. It is easy to change
the rules, without changing the scaling exponent, to obtain a
network that displays nonzero clustering. Indeed, a version in
which each two copies of the original hub 
are connected has a finite clustering coefficient. 
These and further variants of the model
  will be discussed in forthcoming publications.

    A crucial unique property of the model is its hierarchical
structure.
    It is not clear whether the hierarchy seen in this model is a unique property
of its construction, or it is an intrinsic property of all
scale-free networks.
    Uncovering the elements of the hierarchy could be an important future task in our
understanding of scale-free networks.
    The   present model explicitly builds such a hierarchical construction into
the network, as at each iteration we connect nodes to each
  other in a hierarchical fashion.

In conclusion, we introduced a simple model that allows us to construct
a deterministic scale-free network.
    We have shown that the scaling exponent characterizing
the tail of the $P(k)$ distribution can be calculated analytically.
    An important feature of the model is its hierarchical topology,
as at each iteration it combines identical elements to generate a
larger network.
    The method that lead
  to the construction of the network naturally lends itself to immediate generalizations,
  leading to structures that are similar
   in spirit, but rather different in detail,
   allowing us to generate networks with different scaling exponents and connectivity.

{\bf Acknowledgments:}

We with to thank H. Jeong for his help with the manuscript.
This research was supported by NSF-PHY-9988674.

\end{document}